\documentclass[]{emulateapj}

\usepackage{epsfig}
\usepackage{natbib}
\usepackage{graphicx}	
\usepackage{amsmath}
\usepackage{amsfonts}
\usepackage{amssymb}
\usepackage{color}
\usepackage{rotating} 

\def\psrA{PSR~J1119$-$6127}
\def\psrB{PSR~J1718$-$3718}
\def\psrC{PSR~J1734$-$3333}
\def\psrD{PSR~J1846$-$0258}
\def\CTA1{PSR~J0007+7303}

\def\1119{PSR~J1119$-$6127}
\def\J1846{PSR~J1846$-$0258}

\def\fermi{{\em Fermi}}
\def\rxte{{\em Rossi X-ray Timing Explorer}}
\def\chandra{{\em Chandra X-ray Observatory}}
\def\xmm{XMM-{\em Newton}}

\def\degr{\hbox{$^\circ$}}
\def\arcmin{\hbox{$^\prime$}}

\def\fdg{\hbox{$.\!\!^\circ$}}

\def\ecs{erg\,cm$^{-2}$\,s$^{-1}$}
\def\es{erg\,s$^{-1}$}

\def\ergs{\,erg\,s$^{-1}$}

\def\gam{$\gamma$}

\def\index{$1.0 \pm 0.3 ^{+0.4}_{-0.2}$}
\def\cutoff{$0.8 \pm 0.2 ^{+2.0}_{-0.5}$}
\def\flux{$9.3 \pm 1.2 \pm 2.0$}
\def\eflux{$6.4 \pm 0.5 \pm 1.0$}
\def\lum{$54 \pm 6 \pm 8$}
\def\eff{$0.23 \pm 0.03 \pm 0.04$}
\def\deltapeak{$0.43 \pm 0.02$}

\def\efluxoff{$2.0 \pm 0.3$}
\def\indexoff{$0.2 \pm 1.0$}
\def\cutoffoff{$0.3 \pm 0.2$}

\shortauthors{PARENT ET AL.}
\shorttitle{Fermi Observations of high-B pulsars}

\slugcomment{Accepted for publication in ApJ, Draft version \today}

\begin{document}
\bibliographystyle{apj}

\renewcommand{\arraystretch}{0.9}

\title{Observations of energetic high magnetic field pulsars \\ with the Fermi Large Area Telescope}

\author{
D.~Parent\altaffilmark{1}, M.~Kerr\altaffilmark{2}, P.~R.~den~Hartog\altaffilmark{2}, 
M.~G.~Baring\altaffilmark{3}, M.~E.~DeCesar\altaffilmark{4,5}, C.~M.~Espinoza\altaffilmark{6}, 
E.~V.~Gotthelf\altaffilmark{7}, A.~K.~Harding\altaffilmark{4}, S.~Johnston\altaffilmark{11}, 
V.~M.~Kaspi\altaffilmark{14}, M.~Livingstone\altaffilmark{14}, R.~W.~Romani\altaffilmark{2}, 
B.~W.~Stappers\altaffilmark{6}, K.~Watters\altaffilmark{2}, P.~Weltevrede\altaffilmark{6},
A.~A.~Abdo\altaffilmark{1},
M.~Burgay\altaffilmark{13}, 
F.~Camilo\altaffilmark{7}, 
H.~A.~Craig\altaffilmark{2}, 
P.~C.~C.~Freire\altaffilmark{10}, 
F.~Giordano\altaffilmark{8,9}, 
L.~Guillemot\altaffilmark{10}, 
G.~Hobbs\altaffilmark{11}, 
M.~Keith\altaffilmark{11}, 
M.~Kramer\altaffilmark{6,10}, 
A.~G.~Lyne\altaffilmark{6}, 
R.~N.~Manchester\altaffilmark{11}, 
A.~Noutsos\altaffilmark{10}, 
A.~Possenti\altaffilmark{13}, 
D.~A.~Smith\altaffilmark{12} 
}

\altaffiltext{1}{College of Science, George Mason University, Fairfax, VA 22030, resident at Naval Research Laboratory, Washington, DC 20375, USA; dmnparent@gmail.com}
\altaffiltext{2}{W. W. Hansen Experimental Physics Laboratory, Kavli Institute for Particle Astrophysics and Cosmology, Department of Physics and SLAC National Accelerator Laboratory, Stanford University, Stanford, CA 94305, USA; kerrm@stanford.edu; hartog@stanford.edu}
\altaffiltext{3}{Rice University, Department of Physics and Astronomy, MS-108, P. O. Box 1892, Houston, TX 77251, USA}
\altaffiltext{4}{NASA Goddard Space Flight Center, Greenbelt, MD 20771, USA}
\altaffiltext{5}{Department of Physics and Department of Astronomy, University of Maryland, College Park, MD 20742, USA}
\altaffiltext{6}{Jodrell Bank Centre for Astrophysics, School of Physics and Astronomy, The University of Manchester, M13 9PL, UK}
\altaffiltext{7}{Columbia Astrophysics Laboratory, Columbia University, New York, NY 10027, USA}
\altaffiltext{8}{Dipartimento di Fisica ``M. Merlin" dell'Universit\`a e del Politecnico di Bari, I-70126 Bari, Italy}
\altaffiltext{9}{Istituto Nazionale di Fisica Nucleare, Sezione di Bari, 70126 Bari, Italy}
\altaffiltext{10}{Max-Planck-Institut f\"ur Radioastronomie, Auf dem H\"ugel 69, 53121 Bonn, Germany}
\altaffiltext{11}{CSIRO Astronomy and Space Science, Australia Telescope National Facility, PO Box 76, Epping NSW 1710, Australia}
\altaffiltext{12}{Universit\'e Bordeaux 1, CNRS/IN2p3, Centre d'\'Etudes Nucl\'eaires de Bordeaux Gradignan, 33175 Gradignan, France}
\altaffiltext{13}{INAF - Cagliari Astronomical Observatory, I-09012 Capoterra (CA), Italy}
\altaffiltext{14}{Department of Physics, McGill University, Montreal, PQ, Canada H3A 2T8}


\begin{abstract}

We report the detection of \gam-ray pulsations from the high-magnetic-field rotation-powered pulsar PSR~J1119$-$6127 using data from the {\it Fermi} Large Area Telescope.  The \gam-ray light curve of PSR~J1119$-$6127 shows a single, wide peak offset from the radio peak by \deltapeak\ in phase.  Spectral analysis suggests a power law of index \index\ with an energy cut-off at \cutoff\,GeV. The first uncertainty is statistical and the second is systematic. We discuss the emission models of PSR~J1119$-$6127 and demonstrate that despite the object's high surface magnetic field---near that of magnetars---the field strength and structure in the $\gamma$-ray emitting zone are apparently similar to those of typical young pulsars. Additionally, we present upper limits on the \gam-ray pulsed emission for the magnetically active PSR~J1846$-$0258 in the supernova remnant Kesteven 75 and two other energetic high-{\it B} pulsars, PSRs~J1718$-$3718 and J1734$-$3333.  We explore possible explanations for the non-detection of these three objects, including peculiarities in their emission geometry.

\end{abstract}


\section{Introduction} \label{sec:intro}

Magnetic fields in neutron stars are thought to originate mainly from the fossil fields of the progenitor stars or from field generation during supernova core collapse \citep[see, e.g.,][]{Spruit08_field}. These strong fields can manifest themselves as `classical' rotation-powered pulsars and magnetars, also known as anomalous X-ray pulsars (AXPs) and soft \gam-ray repeaters (SGRs). While pulsars emit steady, beamed electromagnetic radiation from radio up to high energies ultimately via the loss of rotational kinetic energy ($\dot{E} = 4\pi^2 I \dot{P}/P^3$\,\ergs), magnetars dissipate their extremely high surface magnetic fields ($B_s \sim 10^{14} - 10^{15}$\,G) in luminous X-ray emission---variable on multiple time scales---generally exceeding the power derived from losing rotational kinetic energy \citep[see][for reviews]{Woods06_review,Mereghetti08_review}.

The bulk of pulsar magnetic fields inferred from spin parameters are clustered around $10^{12}$\,G. Nevertheless, a few have estimated fields near or surpassing the quantum-critical field $B_{\rm{cr}} = m_{\rm{e}}^2 c^{3} / e \hbar = 4.4 \times 10^{13}$\,G, at which the electron cyclotron energy is equal to its rest mass. This small class of strong-field pulsars provides an opportunity to constrain emission mechanisms at high energy and to explore the interesting interface between pulsars and magnetars. \citet{Camilo00_1119-1814} discovered PSR~J1814$-$1744, the first pulsar breaking the $B_{\rm{cr}}$ barrier with  $B_s = 5.5 \times 10^{13}$\,G and $\dot{E} = 4.7 \times 10^{32}$\,\ergs, and PSR~J1119$-$6127 with a slightly weaker field. The latter object recently exhibited an unusual glitch recovery during which its radio emission shifted from its typical profile, showing an ``intermittent'' peak and ``Rotating-Radio-Transient-like'' characteristics \citep[see][for RRATs]{McLaughlin06_RRAT}. \citet{Weltevrede10_J1119} demonstrated that this type of behavior could be related to reconfigurations of the magnetic field in the magnetosphere. Moreover, PSR~J1846-0258, the pulsar in the supernova remnant (SNR) Kesteven 75, exhibited the first known magnetic activity in a rotation-powered pulsar with a series of 5 SGR-like X-ray bursts and an X-ray brightening (both pulsed and DC emission) lasting about two months \citep{Gavriil08_kes75bursts}. A large spin-up glitch with an unusual recovery coinciding with the onset of its magnetar-like behavior has also been reported \citep{Kuiper09_kes75,Livingstone10_kes75}. In this context, the connection between high-{\it B} pulsars and magnetars is strengthened by the discovery of a low-{\it B} SGR which shows magnetar-like activity despite its canonical magnetic field \citep{Rea10_lowBSGR}.

In the high energy field for high-{\it B} pulsars, the lower energy spectral cut-off between 10 and 30\,MeV observed for PSR~B1509$-$58 in the Energetic Gamma-Ray Experiment Telescope (EGRET) era was attributed to the higher surface magnetic field \citep{Harding97,Thompson08_egretpsr}. However, the recent discovery of GeV pulsations from the pulsar in the supernova remnant CTA1 with a comparable magnetic field and no radio counterpart showed that the emission mechanism is more complicated \citep{Abdo08_CTA1}. These observations clearly encourage closer examination of these objects to help assess if this class has unique properties from a possible overlap with magnetars.

In this paper we present \gam-ray observations, acquired by the Large Area Telescope (LAT) aboard the \fermi\ Gamma-ray Space Telescope, of energetic high-{\it B} pulsars ($B \gtrsim 4 \times 10^{13}$\,G) for which the spin-down energy loss rate is above $10^{33}$\ergs. We report the detection of pulsed \gam-ray emission from \psrA\ and discuss this detection and possible emission mechanism scenarios. We also present an upper limit on the pulsed emissions of the three other candidates including PSR~J1718$-$3718, PSR~J1734$-$3333, and PSR~J1846$-$0258. Table \ref{tab:param} lists measured and derived parameters for the selected pulsars, ordered by their surface field strengths $B_s$.

\psrA\ was discovered in the Parkes multibeam pulsar survey with a period of 408\,ms and a large period derivative of $4.0 \times 10^{-12}$\,s\,s$^{-1}$ \citep{Camilo00_1119-1814}. Subsequently, it was detected  by \xmm\ in the 0.5 - 2.0\,keV range showing a single, narrow pulse aligned with the radio peak \citep{Gonzalez05_J1119}. Besides its large surface dipole magnetic field $B_{s}$ of $4.1 \times 10^{13}$\,G, the measurement of timing parameters infers a very young pulsar ($\tau_c = P/2\dot{P}$) of 1.6\,kyr with a high spin-down power of $\dot{E} = 2.3 \times 10^{36}$ \ergs.  PSR~J1119$-$6127 is associated with the $\sim$15\arcmin\ diameter shell-type SNR G292.2$-$0.5 observed in both radio and X-ray bands \citep{Crawford01_1119,Pivovaroff01_1119}, as well as a compact and faint X-ray pulsar wind nebula (PWN) observed by \chandra\ \citep{Gonzalez03_1119, Safi-Harb08_1119} which is spatially coincident with the TeV source HESS~J1119$-$614 \citep{arache09}. The SNR/\psrA\ system is located in the Galactic plane $8.4 \pm 0.4$\,kpc away, based on neutral hydrogen absorption of X-rays from the SNR \citep{Caswell04_snr_dist}. However, using the NE2001 model of the Galactic electron distribution \citep{Cordes02_NE2001}, we derived a distance of $16.7_{-7.1}^{+\infty}$\,kpc, which places the pulsar beyond the Sagittarius arm. This discrepancy is probably due to the line-of-sight ($l=292.151$, $b=-0.537$), which is nearly aligned with Sagittarius arm, making distance determination more uncertain than usual. In this work, we use the distance $d = 8.4$\,kpc. 

PSR~J1718--3718 is a young pulsar with a large spin period of 3.3\,s and a period derivative of $\dot{P} = 1.6 \times 10^{-12}$\,s\,s$^{-1}$, implying a spin-down luminosity of $1.6 \times 10^{33}$\ergs, the lowest of the four selected pulsars. However, it has the second highest surface magnetic field strength ($B_{S} = 7.4 \times 10^{13}$\,G) among the known radio pulsars. It was discovered in the Parkes Multibeam Survey \citep{Hobbs04_survey} not far from the direction of the Galactic Center ($l=349.85$, $b=0.22$), and observed as a point-like source by \chandra\ with a thermal spectrum resembling that of the transient AXP XTE J1810$-$197 in quiescence \citep{Kaspi05_1718}. Recently, deeper \chandra\ observations have shown pulsations in the soft X-ray band (0.8 - 2\,keV) and better constrain the source spectrum \citep{Zhu10_J1718}. Its distance based on a dispersion measure (DM) of 373\,pc\,cm$^{-3}$ is $d = 4.5 \pm 0.5$\,kpc.

PSR~J1734--3333 is a radio pulsar found in the Parkes Multibeam Survey \citep{Morris02_parkes} with $B_{s} = 5.2 \times 10^{13}$\,G. \citet{Olausen10_1734} report the probable X-ray detection of the pulsar using \xmm\ observations. The NE2001 model assigns a distance of $6.1_{-1.0}^{+1.6}$\,kpc which, with the projected position ($l=354.82$, $b=-0.43$), implies the pulsar is located in the Near 3\,kpc Arm toward the Galactic Center.

PSR~J1846--0258 was discovered in the \rxte\ data \citep[RXTE;][]{Gotthelf00_kes75} with $B_{s} = 4.9 \times 10^{13}$\,G. Its spin period of 324\,ms and its spin-down rate of $\dot{P} = 7.1 \times 10^{-12}$\,s\,s$^{-1}$ indicate this pulsar is one of the youngest known to date ($\tau_c = 723$\,yr). No radio pulsed emission has been detected yet despite deep searches \citep{Archibald08_kes75}. The distance of the associated SNR Kesteven 75 and the pulsar is still in debate. Published estimates have ranged from 6.6\,kpc to 21\,kpc (Caswell et al. 1975, Milne 1979, Becker \& Helfand 1984). A more recent study by \citet{Leahy08_kes75} estimates the distance between 5.1\,kpc and 7.5\,kpc based on HI and $^{13}$CO velocity measurements, while \citet{Su09_kes75} in an detailed study of molecular-cloud velocities from $^{12}$CO measurements determine the distance as $10.6^{+0.1}_{-1.0}$\,kpc. Based on these references, we adopt in this work the distance of $7.9 \pm 2.8$\,kpc.

\section{Observations}\label{sec:obs} 

\subsection{Timing Observations}\label{sec:timingobs}

Among the pulsars considered here, three have spin-down powers above $10^{34}$\ergs (see Table \ref{tab:param}) and therefore have been followed by several observatories in the context of the pulsar timing campaign for \fermi, which monitors the most energetic radio and X-ray pulsars \citep{Smith08_timing}. This extensive program, begun in early 2007 and lasting through the \fermi\ LAT mission, provides accurate measurements of the pulsar rotation parameters used to assign phases to \gam-ray photons. 

PSRs~J1119$-$6127 and J1734$-$3333 are observed approximately monthly with the 64-m Parkes radio telescope in Australia, at a frequency of 1.4\,GHz (and occasional observations at 0.7 and 3.1\,GHz). The Jodrell Bank Observatory also monitors PSR~J1734$-$3333 on a weekly basis with the 76-m Lovell telescope, using a 64\,MHz band centered at 1404\,MHz connected to an analog filterbank. For these pulsars a total of 94 and 78 pulse times of arrival (TOAs) were recorded from mid-2008 to 2010, overlapping the \fermi\ LAT observations. The TEMPO2 timing package \citep{Hobbs06_tempo2} was used to build the timing solutions from the TOAs, which have been fitted to the pulsar rotation frequencies and their derivatives. For PSR~J1119--6127, the fit includes 16 harmonically related sinusoids, using the ``FITWAVES'' option in the TEMPO2 package, to flatten the timing noise. The post-fit rms is 2.1 and 30.7\,ms, or 0.5\% and 2.6\% of the pulsar phase, for PSRs~J1119--6127 and J1734$-$3333 respectively. Full details of the observing and data analysis can be found in \citet{Weltevrede2010} and \citet{Hobbs04_survey}. 

PSR~J1718--3718, due to its relatively low spin-down power, is not a target for the \fermi\ campaign. We used a timing solution constructed as outlined in \citet{Manchester10_J1718}. Timing observations were made between 2009 January 24 and 2009 June 19 using the Parkes radio telescope with the 10-cm receiver, which has a 1024-MHz bandwidth centered at 3100\,MHz, and the Parkes digital filterbank system. The resulting timing solution based on 23 TOAs has a post-fit rms of 3.4\,ms. 

Finally, X-ray timing observations for PSR~J1846--0258 were made weekly using the Proportional Counter Array \citep[PCA;][]{jsg+96} on {\it RXTE}, and were collected in ``Good Xenon'' mode. Photons arriving between 2008 June 5 and 2010 April 6 were extracted from the first xenon detection layer in the energy range 2 - 20\,keV, and folded with the previously published ephemeris \citep{Livingstone10_kes75}. The resulting pulse profiles were cross-correlated with a high significance template, producing a single TOA for each observation. The TOAs were fitted to a timing model using the pulsar timing software package TEMPO\footnote{http://www.atnf.csiro.au/research/pulsar/tempo}. Further details of the analysis can be found in \citet{Livingstone06_kes75} and \citet{Livingstone10_kes75}. Note that the detailed description of the timing analysis and results for PSR~J1119$-$6127 using the PCA instrument are given in \S\ref{sec:xray}.

The timing parameters used in this work will be made available from the \fermi\ Science Support Center\footnote{http://fermi.gsfc.nasa.gov/ssc/data/access/lat/ephems}.

\subsection{Gamma-ray Observations}\label{sec:gamobs}

The LAT aboard \fermi\ is an electron-positron pair conversion telescope and went into orbit on 2008 June 11 \citep{Atwood09_LAT,Abdo09_LATcal}. The telescope covers the 20 MeV to $>300$ GeV energy range with good sensitivity and localization performance (an effective area $\sim$\,8000\,cm$^{2}$ on-axis above 1\,GeV and an angular resolution $\theta_{68} \sim 0.6^\circ$ at 1\,GeV for events in the front section of the tracker). The LAT timing is derived from a GPS clock on the spacecraft, and \gam-rays are hardware time-stamped to an accuracy significantly better than 1\,$\mu$s \citep{Abdo09_LATcal}.

\section{Analysis and Results}

\subsection{Data Selection}\label{sec:select}

For each pulsar, we selected LAT data collected between 2008 August 4 (MJD 54682) when \textit{Fermi} began scanning-mode operations\footnote{Except for J1718--3718 for which the timing solution starts at 54855 MJD.} and the end of the ephemeris validity range (17--29 months). We used ``diffuse'' class events (highest probability of being \gam-ray photons) under the P6\_V3 instrument response function (IRFs), and excluded events with zenith angles $>100^\circ$ to reject atmospheric \gam-rays from the Earth's limb. In addition, we also have excluded for the spectral analysis time intervals when the region of interest (ROI) intersects the Earth's limb. The events were analyzed using the standard software package \textit{Science Tools-09-21-00}\footnote{http://fermi.gsfc.nasa.gov/ssc/data/analysis/scitools/overview.html} (ST) for the \textit{Fermi} LAT data analysis and photon phases were calculated using the ``\fermi-plugin'' available in the TEMPO2 pulsar timing software \citep{rkp+10}.

Multiple scattering at low photon energy dominates the LAT's angular resolution. To approximate as much as possible the instrument \emph{Point Spread Function} (PSF), we selected events for the light curves with an energy-dependent angular radius cut centered on the studied pulsar. This selection is given by $<\theta_{68}(E)> = [(5.12^\circ)^{2} \times (100\,{\rm MeV}/E)^{1.6} + (0.07^\circ)^{2}]^{1/2}$ 
which approximates a 68\% containment angle according to the IRFs. 

We searched for \gam-ray pulsations using the bin-independent H-test \citep{deJager10_Htest}, and selecting photons passing the energy-dependent cut for different maximum angular radii (from 0.5 to 5 degrees) around the pulsar position and energy bands. This truncates the point-spread function at low energies and decreases the number of background events. After taking the number of trials into account, we detected a pulsed \gam-ray signal from PSR~J1119$-$6127 with a significance above 5\,$\sigma$ (our detection threshold). For the other pulsars we found significances below 3\,$\sigma$ as discussed in \S~\ref{sec:upperlim}.

\subsection{Gamma-ray Study of \psrA} \label{sec:ana_j1119}

\subsubsection{Light Curves}

\begin{figure}[ht]
\begin{center}
\includegraphics[scale=0.43]{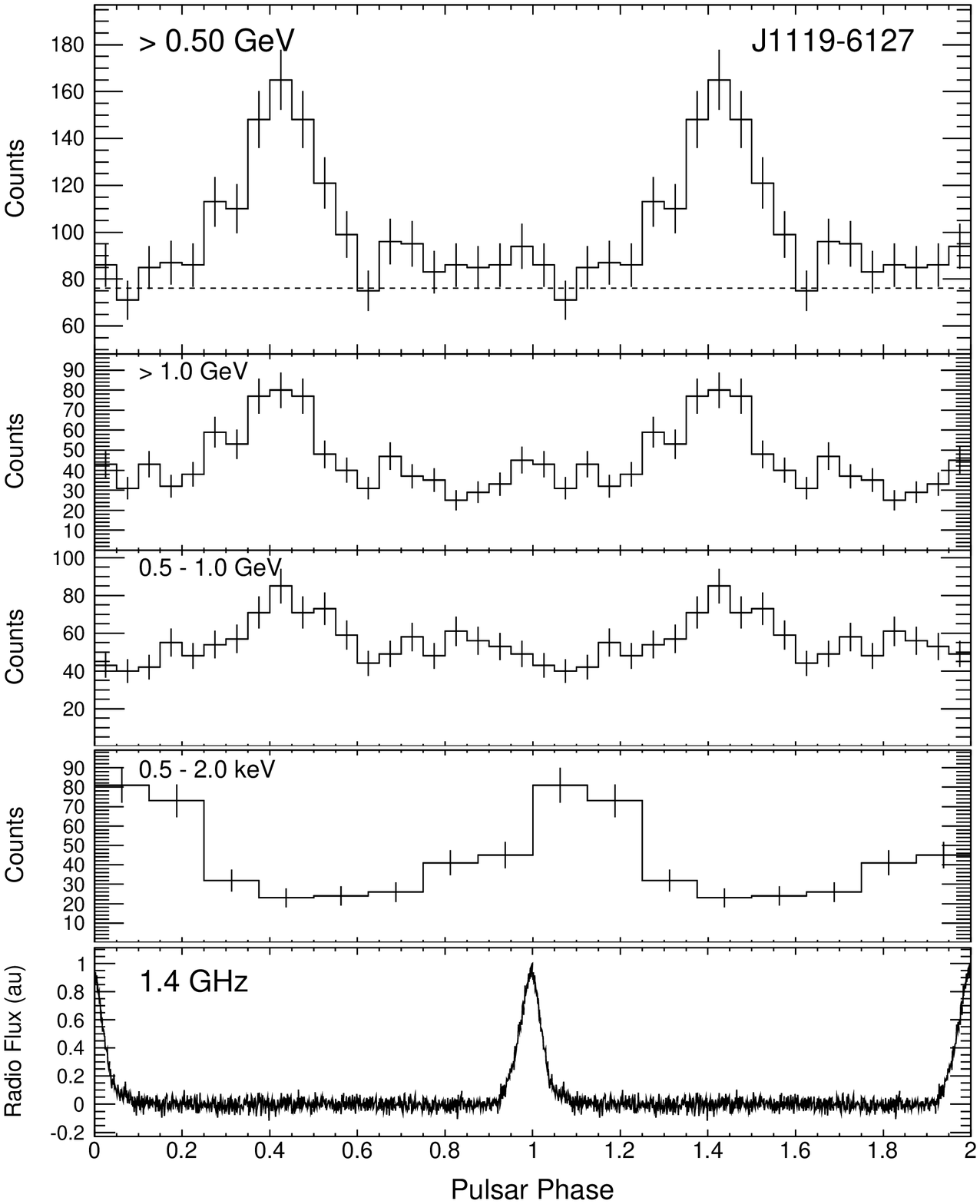}
\caption{Phase-aligned radio, X-ray, and \gam-ray profiles of \psrA. Bottom panel: 1.4\,GHz radio profile observed with the Parkes telescope. Second panel from bottom: X-ray profile in the 0.5--2.0 keV ranges observed by \textit{XMM} \citep[adapted from][]{Gonzalez05_J1119}. Top panels: \gam-ray profiles obtained with the \fermi\ Large Area Telescope in different energy bands. Two rotations are shown. The dashed line shows the background level estimated from a surrounding ring. \label{fig:lc_j1119}}
\end{center}
\end{figure}

Figure \ref{fig:lc_j1119} presents the 20 bin \gam-ray phase histograms in three energy ranges, along with the radio profile and the X-ray profile observed by \xmm\ (bottom panels). The top panel shows the profile with optimized signal-to-noise ratio, corresponding to an event selection above 0.5\,GeV within $0\fdg5$ of the timing position and passing the energy-dependent cut. For this energy band, the H-test gives a value of 107 corresponding to a pulsation significance of $9\,\sigma$.  The dashed line shows the background level (76 counts per bin) estimated using an annular ring centered on the radio position, during the off-pulse window ($\phi < 0.10$ and $\phi > 0.65$), and with inner and outer radii of $0\fdg5$ and $1\fdg5$, respectively. No significant pulsed signal was detected below 0.5\,GeV, despite searches using different apertures. The light curve consists of one single, wide peak observed between 0.10 and 0.65 in phase, which defines our on-pulse window. The light curves between $0.5 - 1$\, GeV and above 1\,GeV within the same aperture have similar pulse profiles. We fitted the unbinned \gam-ray data above 0.5\,GeV with a Gaussian function. The peak is offset in phase from the radio peak by $\delta =$ \deltapeak\ with a FWHM of $0.18 \pm 0.03$. The bias due to the DM uncertainty in extrapolating the radio TOA to infinite frequency is negligible.


\subsubsection{Spectrum}\label{sec:spec_j1119}

\begin{figure}[ht]
\begin{center}
\includegraphics[scale=0.26]{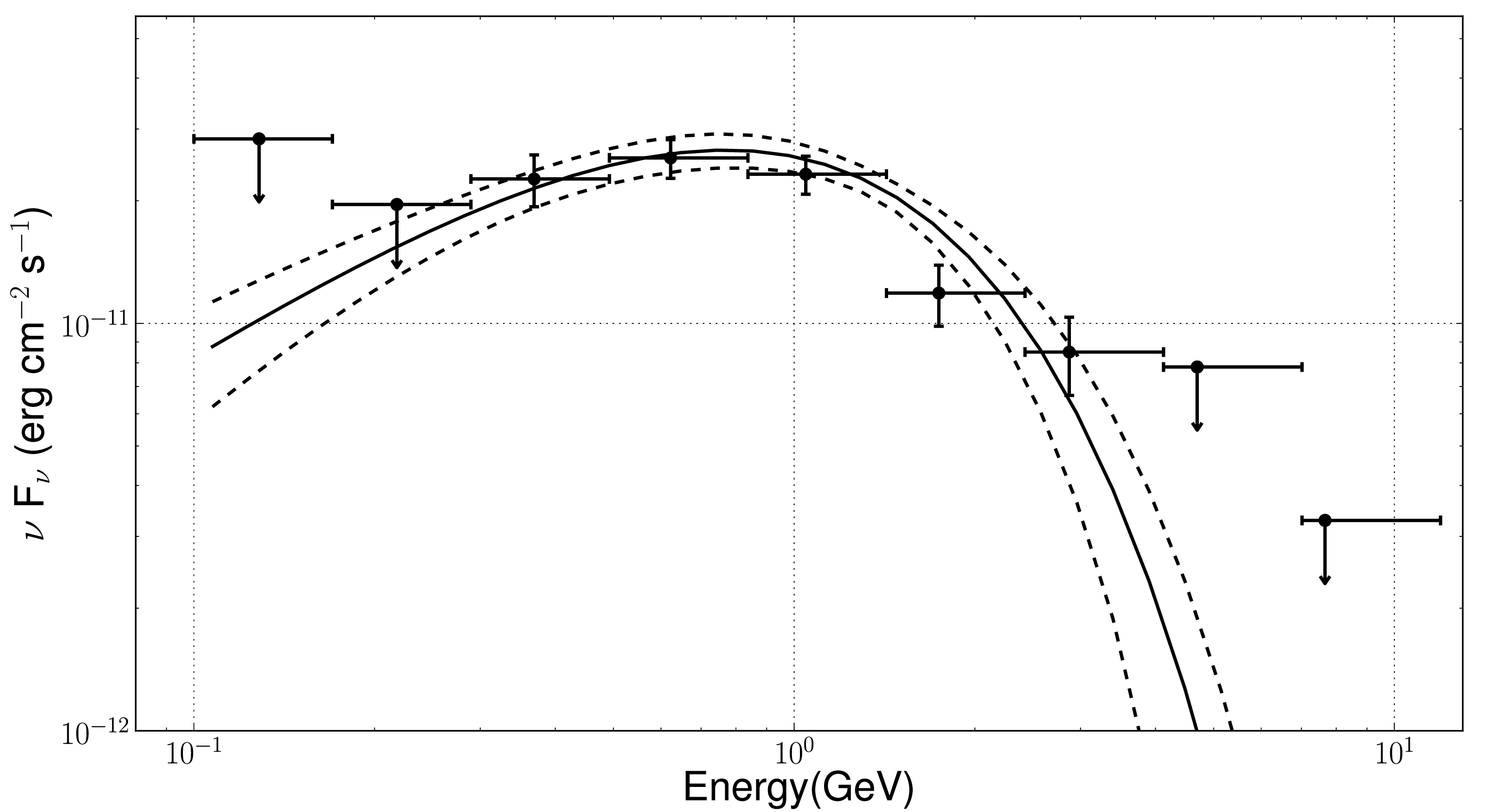}
\caption{Phase-averaged spectral energy distribution of PSR~J1119$-$6127 in \gam-rays obtained with the \fermi\ Large Area Telescope. Plotted points are from likelihood fits to individual energy bands with $\geq 4$\,$\sigma$ detection above background for two degrees of freedom, otherwise an upper limit arrow is shown. The solid black line shows the maximum likelihood fit to a power law with exponential cut-off (Eq. \ref{eq:spectra}). The dashed lines are $\pm$ 1 $\sigma$ uncertainties on the fit parameters. \label{fig:sed_j1119}}
\end{center}
\end{figure}

A phase-averaged spectrum was obtained with a binned likelihood analysis \citep{Mattox96_likelyhood} of the LAT data selected from $20^{\circ} \times 20^{\circ}$ region centered on the pulsar position, using the \fermi\ Science Tool ``gtlike''. The Galactic diffuse emission was modeled using the \textit{gll\_iem\_v02} map cube, while the extragalactic emission and residual instrument backgrounds were modeled jointly by the isotropic component \textit{isotropic\_iem\_v02}. These two models and an explanation of their development are available from the FSSC. In addition, all point sources within $15^\circ$ of the pulsar found in an updated version of the LAT 1FGL source catalog \citep{Abdo10_1FLcat}, using 18 months of data, were included in the model. Sources were modeled with a power law spectrum, except for pulsars, for which a power law with an exponential cut-off was used \citep[see Eq.\ref{eq:spectra} and][]{Abdo10_PSRcat}. Sources more than 5$^\circ$ from the pulsar were assigned fixed spectra taken from the source catalog. Spectral parameters for sources within 5$^\circ$ of the pulsar were left free for the analysis. We fitted the spectrum of PSR~J1119$-$6127 above 100\,MeV using an exponentially cut-off power-law of the form 

\begin{equation}\label{eq:spectra}
{dN \over dE} = N_0 E^{-\Gamma} {\rm exp} \left[-\left(\frac{E}{E_{c}}\right)^{\beta}\right]\, {\rm cm}^{-2}\, {\rm s}^{-1}\, {\rm MeV}^{-1},
\end{equation}
where $N_0$ is the differential flux (in units of ph\,cm$^{-2}$\,s$^{-1}$\,MeV$^{-1}$), $\Gamma$ the photon index, and $E_{c}$ the cut-off energy. The parameter $\beta$, which determines the steepness of the exponential cut-off, was fixed to 1. The \fermi\ LAT pulsars are generally well-described by a simple exponential model, $\beta=1$, a shape predicted by outer magnetosphere emission models \citep{Abdo10_PSRcat}. The energy at which the normalization factor $N_0$ is defined is 1\,GeV. The best-fit values are listed in Table \ref{tab:param} as well as both the photon flux $F_{100}$ and energy flux $G_{100}$ above 100 MeV, where the first errors are statistical and the second are systematic. The systematic uncertainties were estimated by applying the same fitting procedures described above and comparing results using bracketing IRFs where the effective area has been perturbed by $\pm$ 10\% at 0.1 GeV, $\pm$ 5\% near 0.5 GeV, and $\pm$ 20\% at 10 GeV with linear interpolations in the logarithm of the energy for intermediate energies. We also modeled the pulsar with a simple power-law spectrum ($\beta = 0$), and by leaving the $\beta$ parameter free in Equation \ref{eq:spectra}. In the first case, the exponentially cut-off power law model is preferred at the 6\,$\sigma$ level according to the likelihood ratio test \citep{Mattox96_likelyhood}, while in the second case the extra free parameter did not improve the quality of the fit. The results were all cross-checked using an alternate analysis tool developed by the LAT team. Figure \ref{fig:sed_j1119} shows both the phase-averaged spectral fit between 0.1 and 10\,GeV (solid lines) with $\beta = 1$, and the spectral points derived from likelihood fits to each individual energy band in which it was assumed the pulsar had a power-law spectrum.

To search for unpulsed emission from a compact nebula or magnetospheric emission from the pulsar, we searched in the off-pulse region ($\phi < 0.10$ and $\phi > 0.65$) for a point source in the energy band $0.1 - 100$\,GeV at the radio pulsar position. Using a power-law spectrum we found a 7\,$\sigma$ signal with an energy flux above 0.1\,GeV of (\efluxoff)$\times 10^{-11}$\,\ergs\ which represents $\sim 30$\% of the phase-averaged emission. There are three possibilities to explain this signal: emission from a PWN or SNR, magnetospheric pulsar emission, or unmodeled structure in the Galactic diffuse emission. To determine its origin, we fitted the signal in three energy bands (0.1 - 1\,GeV, 1 - 10\,GeV, 10 - 100 GeV) and in the whole energy range (0.1 - 100\,GeV) using a pulsar shape spectrum (see eq. \ref{eq:spectra}). The signal is only significant below 1\,GeV and the exponentially cut-off power law model is preferred over a simple power-law model at the 5\,$\sigma$ level, with a spectral index of \indexoff\ and an energy cut-off of \cutoffoff\,GeV. An exponentially cut-off spectrum, which has been observed in the off-pulse of a few pulsars \citep{LAT_PWN2011}, suggests a magnetospheric origin. However, the model for the Galactic diffuse emission is imprecise in this complex region, and such features may also be associated with unmodeled structure in the Galactic diffuse emission. In particular, the apparent low energy cut-off may be a result of the large PSF at low energies ($\theta_{68} \sim 3^\circ$ at 100\,MeV for events in the front section of the tracker). If the emission is in fact an inadequately-modeled diffuse component, the phase-averaged $\gamma$-ray flux for PSR~J1119$-$6127 would be reduced by 30\%. However, because the emission is only present in a narrow band at low energy, we expect the phase-averaged spectral shape to remain largely unchanged.


Finally, the pulsar associated to the source 1FGL~J1119.4$-$6127c does not show any variability in the 1FGL catalog and an internal 18-month LAT source list \citep{Abdo10_1FLcat}. But in view of the recent evidence for changes in the magnetospheric configuration of PSR~J1119$-$6127, a careful search for changes in the spectrum and pulse shape is warranted if a glitch occurs.

\subsection{X-ray Study of \psrA} \label{sec:xray}

We also searched for pulsed X-ray emission from PSR~J1119$-$6127 to compare to our \fermi\ results. The pulsar was monitored regularly as part of an {\it RXTE} monitoring campaign\footnote{PI: F.~P.~Gavriil} to look for magnetar-like behavior as discovered from PSR~J1846$-$0258 \citep{Gavriil08_kes75bursts}. Data were collected using the PCA, an array of five collimated xenon/methane multi-anode proportional counter units (PCUs) sensitive to incoming photons in the 2 - 60\,keV energy range, with a total effective area of approximately 6500\,cm$^2$ and a field of view of $\sim 1^{\circ}$ FWHM. We used all archival XTE data available in HEASARC public archive\footnote{http://heasarc.gsfc.nasa.gov/docs/archive.html} on the pulsar that spanned the valid interval time of the radio ephemeris. All data were taken in high time resolution Good-Xenon mode ($>$100\,$\mu s$). These data sets were processed using the standard reduction and analysis tools. After extracting time intervals contaminated with PCA breakdown events, the data were merged into a single barycenter corrected FITS photon file. We selected counts from the first layer and energy range up to 20 keV and folded the data with the same ephemeris used to analyze the \fermi\ LAT data (see $\S$3.1). No significant signal was found in the XTE data range (MJD 54890 - 55170). This is not surprising since the X-ray pulsed emission detected by \citet{Gonzalez05_J1119} shows strong energy dependence and appears to be confined to the soft X-ray band below 2\,keV. We then folded data from energy bands restricted to the lowest energy PI channels to test all cumulative channels up to 5\,keV. However, again no signal was detected.

\subsection{Emission Models of PSR~J1119$-$6127}\label{sec:em_model}

The first year of pulsar studies with \fermi\ has shown that the $\gamma$-ray emission mechanism for young pulsars is likely to be outer magnetospheric in origin \citep{Abdo10_PSRcat}. Such emission is predicted by several different models. The Outer Gap models \citep[OG,][]{Cheng86_og1, Cheng86_og2, Romani96_og} place the emitting region on a set of open zone field lines paralleling the closed zone boundary; particles accelerate and radiate between the `null charge' surface and the light cylinder, so that only one magnetic pole is visible in each hemisphere and the hollow cone produces the typical double \gam-ray pulse. An alternative picture is the two-pole caustic model \citep[TPC,][]{Dyks03_tpc} which uses a similar set of field lines, but posits acceleration starting near the star surface; both poles are visible in both hemispheres and a double pulse is formed by truncating the emission zone at $\sim 0.8\,r_{LC}$ so that only the trailing edges of the two cones are present in the light curve. More physical pictures, e.g. the slot-gap model \citep[SG,][]{Muslimov04_slotgap}, may have radiation from a wide range of altitudes and so may be approximated by either of these two geometries.

\begin{figure*}[!ht]
\begin{center}
\includegraphics[scale=0.8]{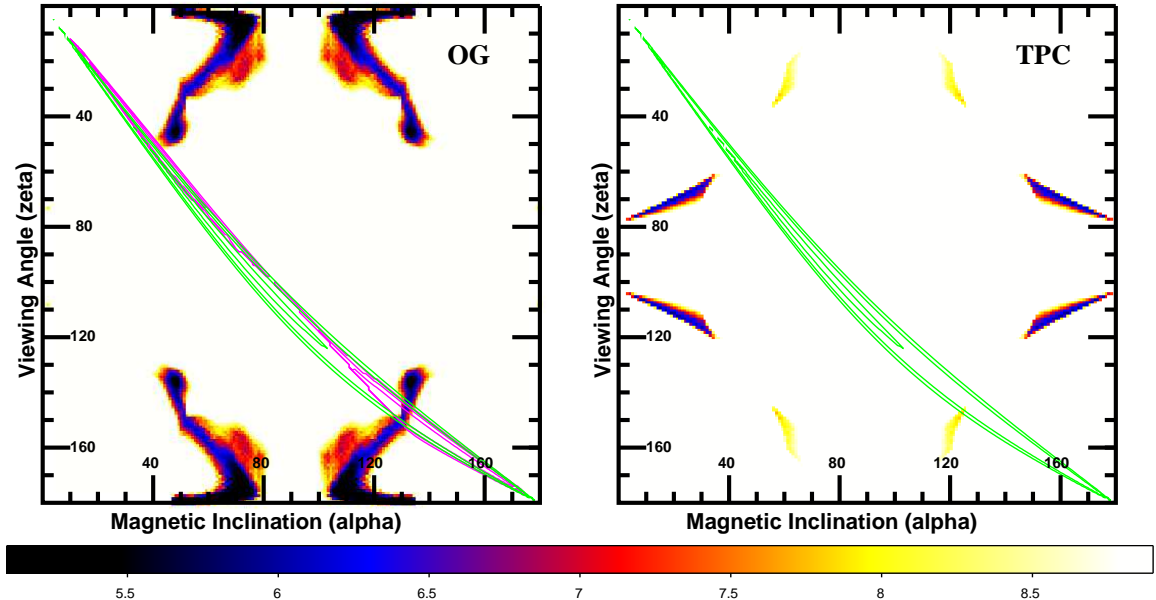}
\caption{Pulsar geometry and emission modeling for PSR~J1119$-$6127 for the `Outer Gap' (left panel) and the `Two Pole Caustic' (right panel) models in the magnetic inclination $-$ viewing angle ($\alpha$, $\zeta$) plane. Green contours show the RVM fit to the \citet{Weltevrede10_J1119} radio polarization data. For the left panel we also show the polarization fit for finite altitude, open zone radio emission ($r=0.09\,r_{LC}$, magenta contours). Contours are at 1.5$\times$, 2.5$\times$, and 3.5$\times$ the minimum value of the reduced $\chi^2 = 0.85$. The background color scale gives the $\chi_3$ statistic fit to the observed $> 500$\,MeV pulse profile. The color scales in the panels are the same, with dark colors representing better fits. Preferred models lie along the diagonal polarization fit band. \label{fig:roger_goodness}}
\end{center}
\end{figure*}

We can compare the predictions of the OG and TPC pictures following the procedure in \citet{Romani10_lcmod}. For these calculations we assume a co-rotating vacuum-like field (`pseudo-force-free'), model the field for a gap width $w = (10^{33} {\rm erg\, s^{-1}}/{\dot E})^{1/2}$ of 0.02 for PSR~J1119$-$6127, and compute the light curve for each magnetic inclination angle $\alpha$ and viewing angle $\zeta$. In Figure~\ref{fig:roger_goodness} we compare the match between the model light curve and the data in the ($\alpha,\, \zeta$) plane using the $\chi_{3}$ weighting defined in \citet{Romani10_lcmod}; dark colors are better fits. For PSR~J1119$-$6127, as for many young \fermi\ pulsars, the radio emission has high linear polarization and has been fit to the rotating vector model (RVM) to constrain the geometrical angles $\alpha$ and $\zeta$ \citep[e.g.,][]{Crawford03_1119}. Even better constraints can be obtained from the unusual double-peaked mode found by \citet{Weltevrede10_J1119} in post-glitch timing observations. We fit the RVM model to these data and correct for the RVM sign convention problem \citep{ew01} to obtain the allowed range (green contours) in Figure~\ref{fig:roger_goodness}. As is often the case, these fits primarily constrain $\beta=\zeta - \alpha$; the $0.5$, $1.5$ and $2.5\,\sigma$ contours from these fits are shown.

\begin{figure}
\begin{center}
\includegraphics[scale=0.44]{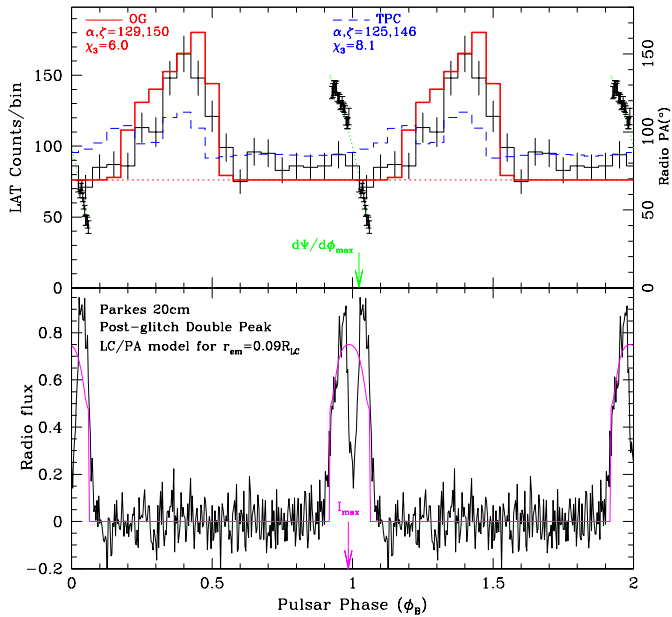}
\caption{Light curves for PSR~J1119$-$6127. The bottom panel shows the Parkes radio light curve in the two peaked (post-glitch) mode, which provides the best model constraints. The corresponding radio polarization position angle data are shown (right scale) in the upper panel. The model Pulsar Phase ($\phi_B$) is referenced to the closest approach of the magnetic axis to the Earth line-of-sight, as fit from the polarization sweep. The sweep rate maximum (green arrow) and pulse profile offsets (magenta arrow) are shown, with good matches to the observed radio data for an altitude of $r=0.09\,r_{LC}$. The upper panel shows the LAT pulse profile (left scale) and model OG (solid line) and TPC (dashed line) profiles. These are best-fit profiles (geometric angles in the legend) and the phase is referenced to the radio-determined phase of the magnetic axis. The $\gamma$-ray background, shown by the dotted line, was estimated using an annular ring centered on the radio position with inner and outer radii of 0.5\degr\ and 1.5\degr\ respectively, during the off-pulse region. \label{fig:roger_lcs} }
\end{center}
\end{figure}

The idealized RVM model assumes a point dipole and does not account for the radio emission altitude or the magnetosphere closed zone. 
If the radio emission arises from finite altitude, the phase and shape of the pulse maximum and the radio polarization sweep are shifted \citep{blask91,dh04}. We have used our numerical models of the magnetospheric field structure to fit the polarization sweep for finite altitude radio emission, requiring that the observed radio flux arise from open field lines. The best fits are found for altitudes $\approx 0.1\,r_{LC}$; we show the contours of the allowed regions in the left (OG) panel of Figure~\ref{fig:roger_goodness}. While the allowed $\alpha$, $\zeta$ range is quite similar to that of the RVM fit, the phase constraints provide important additional information. In Figure~\ref{fig:roger_lcs} we show the observed and model light curves for the radio emission for $\alpha$ and $\zeta$ at the best values allowed by the $\gamma$-ray pulse shape and polarization constraints in Figure~\ref{fig:roger_goodness}. The lower panel shows the Parkes radio light curve in the double-peaked mode. The corresponding positional angle (PA) sweep of the polarization vector is shown in the upper panel (right scale). The phasing is set by the fit (green dotted line) to the maximum rate of the PA sweep, ${\rm d}\Psi/{\rm d}\phi_{\rm max}$ (top panel). This fit is very good for the inferred angles ($\chi^2$/DoF=1.1) and, in turn, determines the phase of the closest approach of the surface magnetic dipole axis to the Earth line-of-sight. This is denoted by the pulsar phase $\phi_B=0$. Note that the phase of the observed pulse centroid (${\rm I_{max}}$, lower panel) is offset from the dipole axis. The intensity and PA offsets are in the same sense as the analytic approximation by \citet{blask91}, but field line sweep back distortion \citep{dh04} makes the offsets smaller, implying a higher emission altitude. Note also that \citet{Weltevrede10_J1119}, using the \citet{blask91} approximations, derived a lower emission height. Thus to fit the observed pulse width into the open zone they argued for small $\alpha$. With the numerical modeling of the polarization sweep and pulse offset we derive larger heights and find that the observed sweep and pulse profile are well matched for angles consistent with the $\gamma$-ray data, while having the radio emission arise from the open zone.

We can now compare with the observed \fermi\ $\gamma$-ray light curve (Figure~\ref{fig:roger_lcs}, upper panel). Adding the constraints supplied by the polarization data restricts the acceptable region. For the OG model the best $\gamma$-ray fits consistent with the polarization constraints are at $\alpha = 125 - 130^\circ$, $\zeta = 140 - 150^\circ$. For the TPC picture best fits in the allowed range are also near $\alpha = 125^\circ$, $\zeta =145^\circ$, however this model gives a poorer pulse profile ($\chi_3 = 8.1$ v.s 6.0) fit than the OG case (Figure~\ref{fig:roger_goodness}). The \fermi\ light curve is referenced to the radio pulse peak (in the single-peaked mode) and the observed radio-peak/$\gamma$-pulse offset is $\delta=0.43$. The $\gamma$-ray pulse models are computed relative to the magnetic axis ($\phi_B = 0$); we find that the best-fit OG light curve (solid line), in fact, has phase $\phi_B = 0.39$ for altitudes of $\approx 0.1\,r_{LC}$, corresponding to $\delta = 0.43$. The model pulse is single, with little off-pulse emission. Using the phase and viewing angles determined by the radio observations, the best-fit TPC model has two peaks and a large unpulsed component (Figure 4). For these models the flux correction factors are $f_\Omega = 1.50$ (OG) and $f_\Omega = 0.95$ (TPC).


Note that these fits assume the background level inferred from the surrounding ring. If, in contrast, the true minimum of the magnetospheric flux is the baseline level at $\phi=0.65-0.1$, then we find that the OG light curves fit much better than the TPC case ($\chi_3=2.0$ vs. 11.3); the best fit $\alpha$ and $\zeta$ are nearly unchanged.

We conclude that, in the context of these simple geometrical models, the high altitude component conventionally associated with the OG model is the preferred origin of the observed $\gamma$-ray pulse, with an excellent match to the observed pulse phase and light curve shape. However, as for other single-peaked $\gamma$-ray pulsars it is difficult to exclude other model scenarios. In particular, some versions of the slot gap models that allow emission at larger altitudes may provide a good fit to the light curve; we will need higher sensitivity to search for the fainter low altitude off-pulse emission predicted to occur in these models. Also, more physical realizations of the pulsar magnetosphere, such as the numerical force-free `Separatrix Layer' (SL) model of \citet{baispit10} may produce acceptable light curve fits. In summary, we find that the light curve of J1119$-$6127 can be well fit by conventional $\gamma$-ray pulsar models but only when emission from relatively close to the light cylinder dominates. Despite the object's high surface magnetic field, the field strength and structure in the $\gamma$-ray emitting zone are apparently similar to those of more typical young pulsars.

\subsection{Gamma-ray Upper Limits for Other High Magnetic Field Pulsars}\label{sec:upperlim}

To calculate an upper limit on pulsed flux, we simulated a point source at the position of the pulsar using the Science Tool \textit{gtobssim} and the \textit{P6\_V3\_DIFFUSE} IRFs.  We used the actual pointing history of the spacecraft to accurately account for the changes in rocking profile during the mission.  Such profile changes affect the exposure a given source receives and can appreciably alter sensitivity. The time intervals of the simulations are limited to the range of validity of the timing solution for each pulsar, an integration time of about 73, 82, and 87 weeks for the pulsars J1718$-$3718, J1734$-$3333, and J1846$-$0258.  Since there is no detection, pulsed or otherwise, of these three pulsars, we adopted a spectral shape for the simulated source typical of the shape observed for other LAT pulsars, namely a power law with exponential cut-off (see Eq. \ref{eq:spectra}), with $\Gamma = 1.5$, $E_c=3$\,GeV, and $\beta = 1$.  We note that higher values of $E_c$ and lower values of $\Gamma$ --- i.e., sources with a greater fraction of their emission at higher energies, where the LAT has superior angular resolution and the Galactic diffuse emission is less intense --- are easier to detect with both pulsed and unpulsed methods.  In addition to the candidate pulsar, we simulated events from the Galactic (\textit{gll\_iem\_v02}) and isotropic (\textit{isotropic\_iem\_v02}) diffuse backgrounds.

Next we modeled the pulsar's light curve as a single Gaussian peak and assigned to each simulated ``pulsar'' photon a phase drawn at random from this light curve.  To the diffuse background photons we assigned a phase drawn at random from a uniform distribution.  To calculate the significance of pulsation, we selected photons from a $5\times5$ grid of circular apertures constructed such that all included photons had a reconstructed position within $\delta\deg$ of the point source and a reconstructed energy above a threshold $E_{th}$, with $\delta$ linearly spaced between $0\fdg5$ and $1\fdg0$, and $E_{th}$ uniformly spaced in logarithmic energy between $100$\,MeV and $1$\,GeV.  For each photon set, we calculated the H-test statistic \citep{deJager10_Htest}, and the pulsed significance was determined as the chance probability of the maximum observed test statistic multiplied by a ``trials factor'' of 25. This combination of pulsation test statistic and grid search yields good sensitivity to pulsations for a source of unknown spectrum and unknown light curve in an arbitrary background.

\begin{figure}
\begin{center}
\includegraphics[scale=0.45]{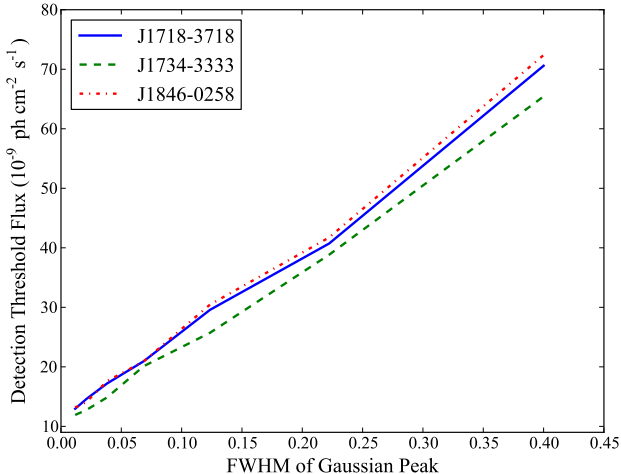}
\caption{The detection threshold for pulsed emission assuming a 100\% pulsed fraction from three pulsars as a function of peak width. The width is characterized by the FWHM of single gaussian peak of the assumed light curve, and the flux reported is that for which 68\% of the simulated sources have a pulsed significance (calculated according to the method described in the main text) greater than $5\sigma$. The integral flux reported is for photon energies above 100\,MeV. \label{fig:pulsed_upper_limits}}
\end{center}
\end{figure}

To determine the upper limit on pulsed emission, we simulated an ensemble of 20 sources at a series of integral fluxes, and for each ensemble we determined the distribution of pulsation significances via the above method.  The upper limit was then simply the flux at which $68\%$ of the sources had a pulsed significance above a given threshold, here taken to be $5\sigma$, or a chance probability less than $5.8\times10^{-7}$.  We note there is an approximately linear relationship between the flux threshold derived and the significance threshold, so the results reported here can be scaled to less stringent detection criteria.  Finally, we perform this exercise for a series of light curves by varying the width of the gaussian peak. The results, shown in Figure \ref{fig:pulsed_upper_limits}, encompass both the sharp peaks typical of $\gamma$-ray pulsar emission and broader sinusoidal peaks.

Finally, we verified the accuracy of the technique by performing a similar exercise with actual LAT data in place of the simulated background.  The resulting limits agree closely with those derived from simulation, indicating the impact of neglecting the contributions of other point sources to the background is unimportant in deriving the pulsed detection thresholds.

\subsection{Gamma-ray Luminosity and Efficiency}

\begin{figure*}[]
\begin{center}
\includegraphics[scale=0.4]{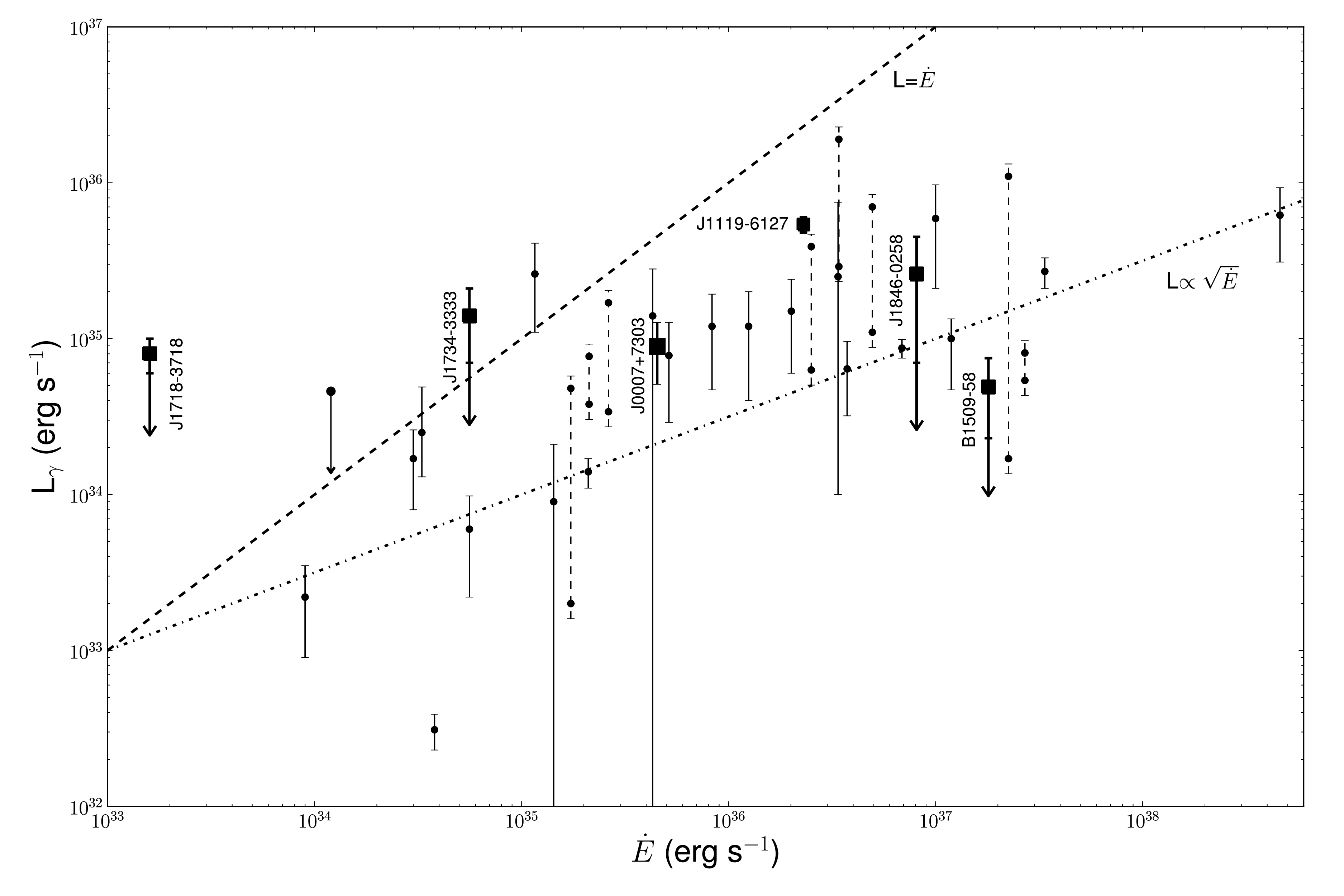}
\caption{Gamma-ray luminosity above 100\,MeV versus the rotational energy loss rate $\dot{E}$. Dashed line: $L_{\gamma}$ equal to $\dot{E}$. Dot-dashed line: $L_{\gamma}$ proportional to the square root of $\dot{E}$. Black points: radio-loud and radio-quiet/faint pulsars. The values are extracted from the first LAT pulsar catalog \citep{Abdo10_PSRcat}, with the addition of PSRs J0248+6021 and J2240+5832 \citep{Theureau10_youngpsrs}. Pulsars with two distance estimates have two markers connected with dashed error bars. The high-{\it B} pulsars are indicated by black squares and the values are reported in Table \ref{tab:param}. \label{fig:lumvsedot}}
\end{center}
\end{figure*}

As argued by \citet{Harding02_polarcap}, pulsars that produce electron-positron pairs through curvature radiation have their primary acceleration limited by the effect of screening of the electric field. This property implies that the high-energy luminosity
\begin{equation}\label{eq:lum}
L_{\gamma} = 4\pi f_{\Omega} G_{100} d^{2} \ {\rm \,erg\,s^{-1}}
\end{equation}
is proportional to $\dot{E}^{1/2}$, and that \gam-ray efficiency $\eta_{\gamma} = L_{\gamma} / \dot{E}$ increases with decreasing spin-down power. In Eq. \ref{eq:lum}, $d$ is the pulsar distance, $f_{\Omega}$ is the geometrical correction factor depending on the inclination and viewing angles $\alpha$, $\zeta$ (see \S \ref{sec:em_model}), and $G_{100}$ is the energy flux measured above 0.1 GeV. Figure \ref{fig:lumvsedot} shows for the studied pulsars, as well as the normal \gam-ray pulsars reported in the first \fermi-LAT Pulsar Catalog \citep{Abdo10_PSRcat} and \citet{Theureau10_youngpsrs}, the $L_{\gamma}$ as a function of $\dot{E}$. The reported values are based on the pulsar distances discussed in this paper, and under the assumption that $f_{\Omega} = 1$ in order to compare the results with the pulsar catalog.

For PSR~J1119$-$6127, summing the distance and $G_{100}$ uncertainties in quadrature yields $L_{\gamma} =$ (\lum)$\times 10^{34} f_{\Omega}$\,\ergs. This places the pulsar between the heuristic constant voltage line $L_{\gamma}^{h} = 10^{33} \times (\dot{E}/10^{33})^{1/2}$\,\ergs\ represented by the dot-dashed line and the line for 100\% conversion efficiency ($L_{\gamma} = \dot{E}$). This value confirms that in the range $10^{35}$\,\ergs\ $< \dot{E} < 10^{36.5}$\,\ergs, $L_{\gamma}$ seems flat. For the emission models discussed in \S\ref{sec:em_model},  the derived flux correction factors are $f_{\Omega} = 1.50$ (OG) and $f_{\Omega} = 0.95$ (TPC), and hence the luminosity cited above may be underestimated for the OG model. 

For PSR~J1846$-$0258, we derived from the upper limit on pulsed flux a luminosity of $(26 \pm 19) \times 10^{34} f_{\Omega}$\,\ergs, using a distance of $7.9 \pm 2.8$\,kpc and assuming a Gaussian peak with a FWHM of 0.2. The upper limit overlaps the $L_{\gamma}^{h}$ line for small distances ($\sim 5$\,kpc), which is more constraining if PSR~J1846$-$0258 behaves as a rotation-powered pulsar. However for large distances in the direction of the far side of the Sagittarius arm, the upper limit is well above the constant line, matching the luminosity of PSR~J1420$-$6048.

For PSRs~J1718$-$3718 and J1734$-$3333, the derived upper limits on the pulsed luminosity are well above 100\% efficiency, and are thus unconstraining. However, these values may be biased by the adopted distances, which are based on the DM. 

\section{Discussion} \label{sec:disc}

The physics close to the surface of a neutron star with a magnetic field comparable to the quantum-critical field ($B_{\rm{cr}} = 4.4 \times 10^{13}$\,G) requires a full QED (quantum electrodynamics) treatment \citep[see][for a review]{Harding06_review}, and we outline a few of the salient features here.  With this strong field, the cyclotron energy ($\hbar\omega_c$) equals the rest mass of the electron and is much larger than the Coulomb energy.  Particles flow primarily along field lines, and particle momentum transverse to the field lines is quantized in Landau levels.  Resonant absorption and emission into/from these states is a primary feature of the radiation spectra.  Most importantly, the strong field enables otherwise forbidden processes such as single photon pair creation $\gamma\to e^+e^-$ and photon splitting $\gamma\to\gamma\gamma$ to proceed, and in fact become quite probable. Intense pair creation is usually invoked as a probable seed for significant radio emission \citep[e.g.][]{Sturrock71}. Yet these two processes can act to suppress the overall yield of created electron-positron pairs, in turn possibly leading to radio quiet high-{\it B} pulsars and magnetars \citep{Baring98_highbpsr}: photon splitting, because it can operate below the $\gamma\to e^+e^-$ threshold of $2m_ec^2$, and pair creation, because when $B\gtrsim 6\times 10^{12}$\,G, it generates pairs in the ground state \citep{BaringHarding01}, thereby inhibiting cascading and subsequent pair creation. The precise level of reduction of the pair yield by these physical mechanisms is contingent upon a number of factors, especially the photon emission/attenuation locale, and is therefore somewhat uncertain, as discussed at length by \citet{BaringHarding01}. Moreover, radio emission from several high-{\it B} pulsars \citep{Camilo00_1119-1814} and magnetars \citep{Camilo06_1810,Camilo07_1547,Levin10_J1622} has been detected, a direct indication that the radio signal is not totally suppressed. Perhaps it originates from a higher altitude where the magnetic field is weaker and pair suppression is limited.

For \gam-ray emission originating near the surface of the neutron star, the influence of the high magnetic field appears as a super-exponential cut-off in the \gam-ray spectrum if one-photon pair-creation attenuation dominates \citep{Daugherty1982em}, while a more gradual cut-off is expected if photon-splitting dominates \citep{Harding97}. These low altitude processes cannot modify \gam-ray emission originating in the outer magnetosphere.  As we argue below, the similarity of the observational characteristics of PSR~J1119$-$6127 to other ``normal'' \gam-ray pulsars implies any connection between the high surface magnetic field and high-altitude \gam-ray emission is weak. This is underpinned by the exponential character of the turnover in Figure \ref{fig:sed_j1119}.  As discussed for other {\it Fermi} LAT pulsars, the maximum energy $\epsilon_{max}$ of the pulsations provides a lower limit to the altitude of $\gamma$-ray emission, since it must lie below the $\gamma $-B magnetic absorption threshold.  Assuming a dipole field, this provides an estimate of the minimum emission altitude $r > (\epsilon_{max} B_{12}/1.76\,\hbox{GeV})^{2/7}P^{-1/7}R_{*}$ \citep[inverting Eq. 1 of][]{baring04}, where $B_{12} = B_s/10^{12}$\,G is the scaled surface magnetic field and $R_{*}$ is the neutron star radius. For PSR ~J1119$-$6127 we observed pulsations up to $\epsilon_{max} \sim$ 4\,GeV, which places a robust lower limit of $r\gtrsim 4.2\,R_{*}$, thereby precluding emission near the stellar surface.

In the following subsections, we first compare the detection of PSR~J1119$-$6127 with the two other high-field \gam-ray pulsars, namely PSR~B1509$-$58 \citep{Abdo10_B1509} and the radio-quiet pulsar J0007+7307 in CTA1 \citep{Abdo08_CTA1}. We then will discuss possible explanations for the non-detection in \gam-rays of PSR~J1846$-$0258. 

\subsection{PSR~J1119$-$6127}

The characteristics of PSR~J1119$-$6127 determined with \fermi\ are remarkably similar to those of the majority of the \fermi-detected young pulsars \citep{Abdo10_PSRcat}. The delay between the radio peak and the \gam-ray peak, the \gam-ray pulse profile, and the GeV spectrum indicate an outer-magnetosphere geometry with a preference for the OG model as shown in the light-curve modeling (see \S \ref{sec:em_model}, Fig.\ref{fig:roger_lcs}). At such high altitudes, the fields are far lower than $B_{{\rm cr}}$: observe that for PSR~J1119$-$6127, $B_{LC} = 5.7 \times 10^{3}$\,G is somewhat high, but not as extreme as that of other \gam-ray pulsars. This new \gam-ray detection also demonstrates that a pulsar with higher $B_{s}$ than PSR~B1509$-$58 can emit \gam-rays \citep{Abdo10_B1509}, and clearly shows that the emission mechanism is more complex. While PSRs~B1509$-$58 and J1119$-$6127 have similar ages, braking indices, and surface magnetic fields, PSR~B1509$-$58 has a complex light curve with one of the two \gam-peaks slightly preceding the radio peak and a softer \gam-ray spectrum breaking in the MeV band \citep{Kuiper99_1509,Abdo10_B1509}. Neither of these two observed features is fully explained by the high-altitude emission models. Moreover, while PSR~B1509$-$58 shows a non-thermal component up to 10 MeV and turns over in the MeV bands, PSR~J1119$-$6127 presents only thermal emission below 10\,keV but emits GeV pulsations. Comparing PSR~J1119$-$6127 with the CTA1 pulsar, it is interesting to note that apart from similarities in the GeV spectrum (cut-off energy at a few GeV) and soft X-ray pulsations \citep[mainly below 2\,keV,][]{Caraveo10_CTA1}, no radio pulsations for CTA1 were detected in spite of deep searches \citep{Halpern04_cta1}. One likely explanation is that the radio beam does not sweep across our line of sight, and this suggests that the emission geometry ($\alpha,\zeta$) is different than in PSR~J1119$-$6127.

\subsection{PSR~J1846$-$0258 and other high-{\it B} pulsars}

Based on its high $\dot{E} = 8.1 \times 10^{36}$\,\es, hard X-ray emission \citep[e.g.][$\Gamma = 1.20 \pm 0.01$, 20--250 keV]{Kuiper09_kes75}, and its young age, PSR~J1846$-$0258 seems to be a good \gam-ray pulsar candidate. However, we detect no \gam-ray pulsations from this object, and no source from the 1FGL catalog coincides with the pulsar position. A first explanation for the non-detection of PSR~J1846$-$0258 can be simply a matter of its geometry. Torus modeling in the PWN implies a line of sight angle $\zeta = 62\degr \pm 5\degr$ \citep{Ng08_1846}. \citet{Livingstone06_kes75} infer a magnetic inclination $\alpha$ about 10$^{\circ}$ using spin-down parameters applied to the model of an oblique rotator with a current-starved outer magnetosphere \citep{Melatos97_spindown}. With these $\zeta$ and $\alpha$ values it is unlikely that low-altitude radio emission will be observed. Indeed, adopting these angles no $\gamma$-ray emission is expected from the OG, whereas from the TPC model non-pulsed (DC) or broad single-peaked pulsed emission would still be possible \citep{Watters09_atlas}. We stress that these predictions are highly dependent on the correctness of the model by \citet{Melatos97_spindown}.

Another possibility is that the sensitivity of the \fermi\ LAT is insufficient to detect PSR~J1846$-$0258. The pulsar is near the Galactic plane, 30$^{\circ}$ away from the Galactic center where the diffuse \gam-ray background is very strong. According to the upper limit on pulsed flux presented in \S\ref{sec:upperlim} and the distance of $7.9 \pm 2.8$\,kpc, the derived \gam-ray efficiency is less than $0.03 \pm 0.02$. The Vela pulsar, which has a well-constrained distance around 300\,pc, and shows a similar spin-down power ($\dot{E} = 7 \times 10^{36}$\ergs), has a \gam-ray efficiency of $0.01 \pm 0.002$, which is just at the limit of PSR~J1846$-$0258. In summary, the appreciable distance and high background could make the \gam-ray detection of this pulsar difficult.

Finally, we cannot rule out that the \gam-ray spectrum of PSR~J1846$-$0258 resembles the very soft \gam-ray spectrum of PSR~B1509$-$58. In X-rays, PSR~J1846$-$0258 mimics PSR~B1509$-$58 in many ways. They both emit pulsed non-thermal hard-X-ray emission up to $\sim$250 keV ($\sim 10$ MeV for PSR~B1509$-$58) with broad asymmetric single-peak pulse profiles with similar hard spectral shapes ($\Gamma \sim$ 1.2--1.4), and neither shows soft thermal emission below 2\,keV (except during the 2006 outburst of PSR~J1846$-$0258), although such a signal could be swamped by the non-thermal emission. The pulsed flux of PSR~B1509$-$58 is $\sim 2 \times 10^{-10}$ \ecs\ ($<250$ keV), which is about an order of magnitude higher than that of PSR~J1846--0258 \citep{Marsden97_1509, Cusumano01_1509}. If the spectrum of PSR~J1846$-$0258 peaks in the MeV range as in the case of PSR~B1509$-$58, then the sensitivity of \fermi\ would also be insufficient to measure its pulsations.

The upper limits on emission from PSRs~J1718$-$3718 and J1734$-$3333---which could be quiescent magnetars for which we have not yet detected magnetic activity---are not constraining. \citet{Zhang05_OG_HBPSR} studied OG \citep{Zhang04_OG} predictions of energetic high-{\it B} pulsars before the launch of \fermi. Assuming $\alpha = 55\degr$, they predicted all pulsars in our sample to be detectable with the \fermi\ LAT. However, they assumed a sensitivity of $2 \times 10^{-9}$\,ph\,cm$^{-2}$\,s$^{-1}$ ($>$ 100 MeV) for a 2-year all-sky survey, which is not realistic along the Galactic plane where the diffuse background is very high \citep[see Fig.~9,][]{Abdo10_PSRcat}. Note that for PSR~J1119$-$6127 the predicted \gam-ray flux is an order of magnitude below our observations. \\

In conclusion, \fermi\ observations of high-{\it B} pulsars are as yet insufficient to determine whether these objects form a homogeneous class or have both a ``normal'' component and a ``magnetar'' component.  Of the pulsars with fields above $10^{13}$\,G, only the soft spectrum of PSR~B1509$-$58 is anomalous, whereas the other features are compatible with emission from other, lower-field \gam-ray pulsars.  On the other hand, PSR~J1846$-$0258 has shown magnetar-like behavior, but the \fermi\ non-detection is not constraining. Note that the same high-{\it B} field effects that are in place in the magnetospheres of these pulsars might also affect magnetars' emission. The non-detection of any SGR or AXP with the \fermi-LAT is not surprising in the context of our results. Several magnetars have rotational energies and distances comparable to the high-{\it B} pulsars we report here, but no significant \gam-ray emission has been detected yet \citep{Abdo10_magnetar}. The best hope for resolving the issue may lie with future X-ray polarization measurements (such as with {\it GEMS}\,\footnote{http://heasarc.nasa.gov/docs/gems/} or a successor), which can probe the spin and magnetic geometry of these high-{\it B} pulsars. These measurements, together with deeper \fermi\ limits, may constrain or even exclude emission from the outer magnetosphere. If true, this would indicate that some high-{\it B} pulsars are quiet in the $>100$\,MeV \gam-ray band, and thus form a distinct class of objects that resemble magnetars in some of their characteristics.  At present, it must also be emphasized that PSRs~J1718$-$3718, J1734$-$3333 and PSR~J1846$-$0258 are distant and in the general direction of the Galactic center, raising the possibility that it is difficult to isolate their signals from the diffuse \gam-ray background. 

\section{Summary} \label{sec:sum} 

We have presented the detection of PSR~J1119$-$6127, which has currently the highest inferred surface magnetic field of all \fermi-detected pulsars. The folded light curve shows a single peak which arrives with a phase delay of \deltapeak\ after the radio peak. No significant pulsations have been detected below 500\,MeV. The spectrum above 0.1\,GeV can be described with a hard power law with an exponential cut-off ($\Gamma =$\,\index, $E_{{\rm c}} =$\,\cutoff\,GeV), and is similar to other normal \gam-ray pulsars, except for PSR~B1509$-$58. Fits of the pulse profile to emission models implies that the most likely emission geometry is that described by the Outer-Gap model. The best-fit parameter region (within radio-polarization boundary conditions) indicates a line of sight $\zeta = 145\degr$ and a magnetic inclination angle $\alpha \sim 125\degr$. Finally, from all these observations, the pulsar appears similar to a normal young \gam-ray pulsar with no evidence for a magnetar-like behavior. 

For the other three high-{\it B} pulsars in the sample we have presented pulsed-emission upper limits as a function of pulse width. For PSRs~J1718$-$3718 and J1734$-$3333 limits are not low enough to constrain the emission geometry due to the low spin-down power and the distance, while for the excellent candidate PSR~J1846$-$0258, the non-detection may suggest that the pulsar has a peculiar geometry, a low energy cut-off in its spectrum, or the \fermi\ LAT sensitivity is insufficient due to the distance and the high diffuse background.

\acknowledgments

We would like to thank Andrea Caliandro and Nanda Rea for their useful discussions.

The \textit{Fermi} LAT Collaboration acknowledges generous ongoing support from a number of agencies and institutes that have supported both the development and the operation of the LAT as well as scientific data analysis. These include the National Aeronautics and Space Administration and the Department of Energy in the United States, the Commissariat \`a l'Energie Atomique and the Centre National de la Recherche Scientifique / Institut National de Physique Nucl\'eaire et de Physique des Particules in France, the Agenzia Spaziale Italiana and the Istituto Nazionale di Fisica Nucleare in Italy, the Ministry of Education, Culture, Sports, Science and Technology (MEXT), High Energy Accelerator Research Organisation (KEK) and Japan Aerospace Exploration Agency (JAXA) in Japan, and the K.~A. Wallenberg Foundation, the Swedish Research Council and the Swedish National Space Board in Sweden.

The Parkes radio telescope is part of the Australia Telescope which is funded by the Commonwealth of Australia for operation as a National Facility managed by the CSIRO. The Lovell Telescope is owned and operated by the University of Manchester as part of the Jodrell Bank Centre for Astrophysics with support from the Science and Technology Facilities Council of the United Kingdom.

\clearpage

\begin{deluxetable}{lcccccc}
\tabletypesize{\scriptsize}
\tablewidth{0pt}
\tablecaption{\label{tab:param} Measured and derived parameters for high-Magnetic-Field Rotation-Powered Pulsars}
\tablecolumns{7}
\tablehead{
\colhead{Parameters} & \colhead{J1718$-$3718} & \colhead{J1734$-$3333} & \colhead{J1846$-$0258} &  \colhead{J1119$-$6127} & \colhead{B1509$-$58$^{d}$} & \colhead{J0007$+$7303$^{e}$}
}
\startdata
Period, $P$~(s) \dotfill 							&  3.378 & 1.169 & 0.325 & 0.408 & 0.151 & 0.316 \\
Period derivative, $\dot{P}$~($10^{-12}$\,s\,s$^{-1}$)\dotfill  & 1.61 & 2.28 & 7.08 & 4.02 & 1.54 & 0.36 \\
Surface magnetic field, $B_{s}$~(10$^{13}$\,G) \dotfill 		&  7.4 & 5.2 & 4.9 & 4.1 & 1.5 & 1.1 \\
Magnetic field at light cylinder, $B_{LC}~(10^{3}$\,G) \dotfill	& 0.018 & 0.307 & 13.2 & 5.67 & 42.2 & 3.21 \\
Age, $\tau_{c}~(kyr) $ \dotfill 						&  33.5 & 8.1 & 0.7 & 1.6 & 1.6 & 13.9 \\
$\dot{E}$~(10$^{34}$\,\ergs) \dotfill 					& 0.16 & 5.6 & 810 & 230 & 1800 & 45 \\
Distance, $d$~(kpc) \dotfill 						 	&  $4.5 \pm 0.5$ & $6.1 ^{+1.6}_{-1.0}$ & $7.9 \pm 2.8$ & 8.4 $\pm$ 0.4 & 5.2 $\pm$ 1.4 & 1.4 $\pm$ 0.3 \\
Braking index, $n$~\dotfill 							& ... & $1.0 \pm 0.3^{f}$ & $2.65\pm 0.01^{g}$ & $2.684 \pm 0.002^{h}$ & $2.839 \pm 0.003^{i}$ &  ... \\
Timing data span (months) \dotfill					& 17 & 19 & 20 & 29 & ... & ... \\
\\
Radio-\gam\ peak offset, $\delta$ \dotfill 				& ... & ... & ... & \deltapeak & 0.96 $\pm$ 0.01& ... \\
\gam-ray peak multiplicity \dotfill 					& ... & ... & ... & 1 & 2 & 2 \\
\gam-ray peak separation, $\Delta$ \dotfill 				& ... & ... & ... & one peak & $0.37 \pm 0.02$ & 0.23 $\pm$0.01\\ 
\\
Photon flux, $^{a}F_{100}$ (10$^{-8}$ cm$^{-2}$s$^{-1}$) \dotfill 		& $<3.8$ & $<3.6$ & $<4.0$ & \flux & ... & 30.7 $\pm$ 1.3 \\
Energy flux, $^{a}G_{100}$ (10$^{-11}$ erg cm$^{-2}$s$^{-1}$) \dotfill 	& $<3.3$ & $<3.1$ & $<3.5$ & \eflux & $< 1.5$ & 38.2 $\pm$ 1.3\\
Energy cut-off, $E_{c}$ (GeV) \dotfill 							& ... & ... & ... & \cutoff & ... &4.6 $\pm$ 0.4\\
Spectral Index, $\Gamma$ \dotfill 							& ... & ... & ... &  \index & ... & 1.38 $\pm$ 0.05\\
Luminosity, $^{b}L_{\gamma}$ (10$^{34}$ erg\,s$^{-1}$) \dotfill 		& $<8\pm2$  & $<14 \pm 7$ & $<26 \pm 19$ & \lum & $<4.9 \pm 2.6$ & 8.9 $\pm$ 3.8 \\
Efficiency, $^{b}\eta$ \dotfill 								& $< 50\pm10$ & $<2.5 \pm 1.0$ & $<0.03 \pm 0.02$ & \eff & $<0.003 \pm 0.002$ & 0.20 $\pm$ 0.08
\enddata
\tablenotetext{a}{$E>$0.1 GeV}
\tablenotetext{b}{$f_{\Omega}$ is assumed to be 1 which can result in an efficiency $> 1$.}
\tablenotetext{c}{The errors on the upper limits are dominated by the distance uncertainties.}
\tablerefs{$^{d}$\citet{Abdo10_B1509}; $^{e}$\citet{Abdo10_PSRcat}; $^{f}$\citet{Espinoza10_J1734}; $^{g}$\citet{Livingstone06_kes75}; $^{h}$\citet{Weltevrede10_J1119}; $^{i}$\citet{Liv05_B1509} }
\tablecomments{The upper limits on pulsed flux for PSRs~J1718$-$3718, J1734$-$3333, and J1846$-$0258 assume a Gaussian peak with a FWHM of 0.2 for the \gam-ray profile, and an exponentially cut-off power-law spectrum (see Eq. \ref{eq:spectra}) with $\Gamma = 1.5$, $E_{c} = 3$\,GeV, and $\beta = 1$. The upper limit for PSR~B1509$-$58 is extracted from \citet{Abdo10_B1509} using the LAT 0.1-0.3\,GeV upper limit in Fig. 3.}
\end{deluxetable}
\clearpage

\bibliography{literature}

\end{document}